\documentclass[preprint,12pt]{elsarticle}

\usepackage{fullpage,epsf,amssymb,amsthm,amsfonts,amsmath,latexsym,array,graphicx,extarrows,mathtools,mathstyle,mathrsfs,slashed,subcaption,amsbsy}
\usepackage[utf8]{inputenc} 
\biboptions{numbers,sort&compress}

\journal{Journal of Subatomic Particles and Cosmology}

\begin{document}
 
\begin{frontmatter}

\title{Non-perturbative constraints on perturbation theory at finite temperature}

\author[1]{Peter Lowdon}

\author[1,2]{Owe Philipsen}

\affiliation[1]{organization={ITP, Goethe-Universität Frankfurt am Main},
             addressline={Max-von-Laue-Str. 1},
             city={60438 Frankfurt am Main},
             country={Germany}}
             
\affiliation[2]{organization={John von Neumann Institute for Computing (NIC) at GSI},
             addressline={Planckstr. 1},
             city={64291 Darmstadt},
             country={Germany}}

\begin{abstract}
It has long been understood that the inclusion of temperature in the perturbative treatment of quantum field theories leads to complications that are not present at zero temperature. In these proceedings we report on the non-perturbative obstructions that arise, and how these lead to deviations in the predictions of lattice scalar correlation functions in massive $\phi^{4}$ theory. Using the known non-perturbative spectral constraints satisfied by finite-temperature correlation functions we outline why the presence of distinct particle-like excitations could provide a resolution to these issues.
\end{abstract}

\end{frontmatter}

\section{Introduction}
\label{intro}
 
\noindent
In the perturbative treatment of quantum field theories (QFTs) at finite temperature $T$ it is well-known that complications arise, including a worsening of the convergence properties~\cite{Kapusta:2006pm,Bellac:2011kqa,Blaizot:2003tw,Andersen:2004fp,Laine:2016hma}. These complications stem from the fact that when $T>0$ certain diagrams contain sub-components which are highly sensitive to the infrared dynamics of the theory. Potential resolutions to these infrared problems include the use of effective field theory~\cite{Braaten:1994na}, computations of higher-loop diagrams~\cite{Kajantie:2002wa,Gynther:2007bw}, and various reorganisations of the perturbative series such as optimised infinite resummations~\cite{Braaten:1989mz,Karsch:1997gj,Chiku:1998kd,Andersen:2000yj,Andersen:2008bz}, and variational techniques like the two-particle irreducible (2PI) formalism~\cite{Blaizot:2000fc,vanHees:2001ik,VanHees:2001pf,vanHees:2002bv}. These features are often viewed as a purely perturbative feature of finite-temperature QFTs, but they may in fact be a symptom of a more fundamental non-perturbative constraint: the \textit{Narnhofer-Requardt-Thirring} (NRT) theorem~\cite{Narnhofer:1983hp}. The NRT theorem implies that non-trivial scattering states with real dispersion relations $\omega=E(\vec{p})$ do not exist when $T>0$. This means that a finite-temperature perturbative expansion \textit{cannot} be constructed using either free field, or quasi-particle propagators with purely real poles, and hence undermines the consistency of the standard perturbative approaches~\cite{Landsman:1988ta}. \\

\noindent
The implications of the NRT theorem are a consequence of the physical situation at finite temperature that the dissipative effects of the thermal medium are everywhere present, even at large times. Any incoming or outgoing scattering state must take these effects into account, and therefore cannot have purely real dispersion relations. This reflects a primitive constraint on QFTs at finite temperature which is not restricted to infrared regimes, but will also affect systems at low temperatures, or with non-vanishing mass scales. Evidence for this constraint has been found explicitly in scalar QFTs, where it has been shown that the standard perturbative procedure breaks down at some fixed loop order if the propagators used in the expansion have a real dispersion relation~\cite{Weldon:1998bj,Weldon:1998xr,Weldon:2001vt}. This breakdown occurs because the self-energy of the perturbative propagator develops a specific branch point singularity which prohibits the computation of corrections to the propagator pole. Using a non-real pole in the basic field propagator resolves these problems, but it remains an open question as to how this pole structure is determined in specific QFTs~\cite{Landsman:1988ta,Weldon:1998xr,Bros:2001zs}.  \\

\noindent 
In these proceedings we report on a recent work~\cite{Lowdon:2024atn} which looks to quantify the effect that these non-perturbative finite-temperature constraints have by comparing the predictions of lattice perturbation theory with the results from lattice simulations. We will outline the basic ideas behind this approach, and why the results suggest a promising direction in which to resolve the perturbative issues encountered at finite temperature.

\section{Investigation of $\phi^{4}$ theory on the lattice}

\noindent
In order to investigate the effects of the non-perturbative constraints outlined in Sec.~\ref{intro}, in Ref.~\cite{Lowdon:2024atn} we compared the predictions of correlation functions in lattice perturbation theory with the corresponding results from lattice simulations. Since these simulations contain the full dynamics of the theory, this allowed us to assess the importance of non-perturbative effects as a function of temperature and the coupling strength. For this purpose we analysed real $\phi^{4}$ theory, which has the corresponding lattice action:
\begin{align}
S = a^{4}\sum_{x \in \Lambda_{a}} \left[ \frac{1}{2}\sum_{\mu} \Delta_{\mu}^{f}\phi(x)\Delta_{\mu}^{f}\phi(x)+ \frac{m_{0}^{2}}{2}\phi(x)^{2} +\frac{g_{0}}{4!}\phi(x)^{4} \right], 
\label{eq:latact}
\end{align}
where $\Delta_{\mu}^{f}$ is the lattice forward derivative, and $a>0$ is the lattice spacing. To stay clear of infrared divergences we focussed on the symmetric phase, with bare masses $m_{0}^{2}>0$. For simplicity we performed perturbative computations and lattice simulations of the spatial correlator, which is defined as the corresponding sum of lattice two-point functions:   
\begin{align}
C(z;a,N_{s},N_{\tau}) &=  a^{3}\sum_{\tau,x,y}\langle \phi(\tau,\vec{x})\phi(0)\rangle. 
\label{spatialC}
\end{align}  
By performing both lattice simulations and perturbative predictions on a finite $N_{s}^{3}\times N_{\tau}$ lattice with identical bare parameters $(am_{0},g_{0})$ we were able to directly compare these results, and hence any statistically significant difference could only be due to the perturbative approximation itself. \\

\begin{figure}[t!] 
\centering
\includegraphics[width=0.5\textwidth]{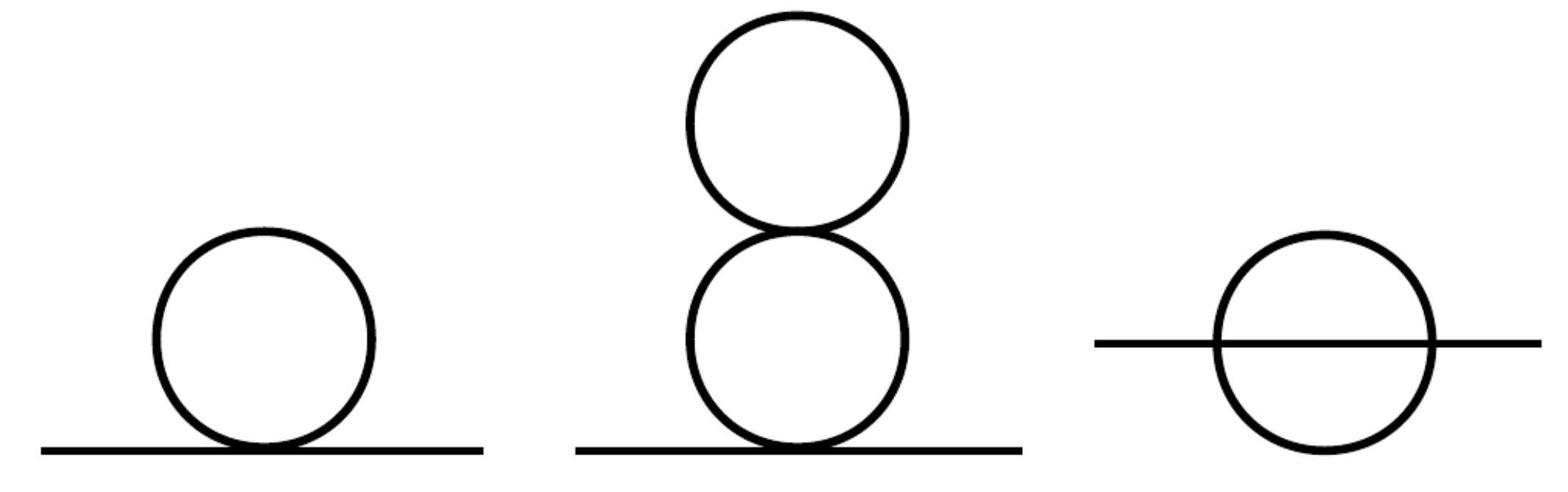}
\caption{The contributing diagrams to the renormalised self-energy up to $\mathcal{O}(g_{0}^{2})$. From left to right these are the tadpole, cactus, and sunset diagrams, respectively.}
\label{self_energy_diag}
\end{figure}

\begin{figure}[t!]
\centering
\includegraphics[width=0.63\textwidth]{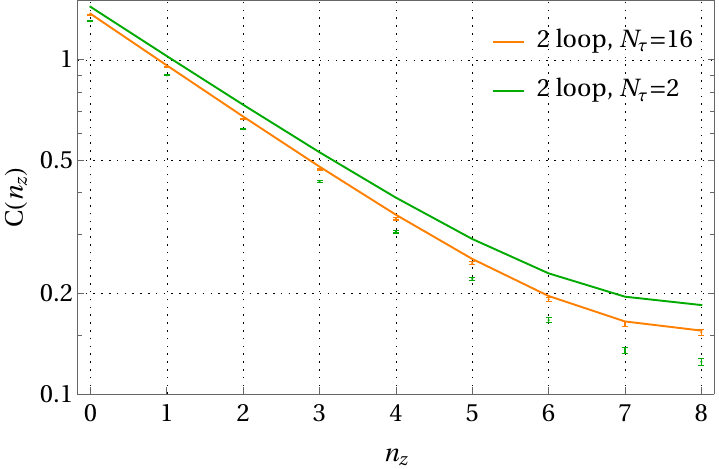}
\caption{Comparison of the two-loop perturbative $C(n_{z})$ predictions with lattice simulations (coloured points) for $N_{s}=16$, $am_{0}=0.15$, and $g_{0}=1.5$. The orange data points ($N_{\tau}=16$) agree well, whereas the green points ($N_{\tau}=2$) display significant deviations from the two-loop predictions.}
\label{fig:L16T16T2}
\end{figure}

\noindent
The lattice perturbation theory predictions are computed in an analogous way as in the continuum, except now any loop integrals are replaced by sums, and the basic fields propagators have the discretised form\footnote{See~\cite{Montvay:1994cy} and references within for a detailed overview of lattice perturbation theory.} 
\begin{align}
\widetilde{G}_{0}(p;a,N_{s},N_{\tau}) = \frac{1}{\sum_{\mu}\tfrac{4}{a^{2}}\sin^{2}\left(\tfrac{a p_{\mu}}{2}\right)+m_{0}^{2}}.
\label{free_prop}
\end{align} 
By defining the self energy $\Pi(p;a,N_{s},N_{\tau})$ of the full propagator to be the sum of all one-particle irreducible diagrams, as in the continuum, it turns out that the spatial correlator in Eq.~\eqref{spatialC} can be written in the following manner:
\begin{align}
C(z;a,N_{s},N_{\tau}) &=   \! \sum_{k_{z}=0}^{N_{s}-1} \! \frac{e^{\frac{2\pi i k_{z}}{aN_{s}}z}}{N_{s}}  \frac{a}{4\sin^{2}\!\left(\tfrac{\pi k_{z} }{N_{s}}\right)+(am_{0})^{2}+a^{2}\Pi(\omega_{E}=p_{x}=p_{y}=0,p_{z}=\frac{2\pi k_{z}}{aN_{s}};a,N_{s},N_{\tau})}.  
\label{Cz}
\end{align}
The spatial correlator can therefore be computed up to some fixed order in $g_{0}$ by first evaluating $\Pi$ at this order, and then performing the sum in Eq.~\eqref{Cz}. In Ref.~\cite{Lowdon:2024atn} we computed predictions up to two-loop order, for which the self-energy has the contributing diagrams in Fig.~\ref{self_energy_diag}. 
Comparing the two-loop perturbative predictions with the lattice simulations, both in the zero-temperature-like case (large $N_{\tau}$) and at finite temperature (small $N_{\tau}$), we found that for sufficiently non-negligible interactions the perturbative predictions increasingly deteriorated as the temperature of the system increased. An example of these deviations are shown in Fig.~\ref{fig:L16T16T2}. In this case the two-loop predictions at the highest temperatures ($N_{\tau}=2$) are actually \textit{larger} than those at one loop. The higher-order corrections are therefore driving the predictions further away from the data such that even the temperature ordering of the correlators is incorrect. This indicates a qualitative departure from the asymptotic-like behaviour of the series in vacuum, and hence a breakdown of the perturbative prediction. \\ 
 
\noindent
The structure of the spatial correlator in Eq.~\eqref{Cz} implies that the deviations in the perturbative predictions from the lattice data must arise from the properties of the self-energy $\Pi(p;a,N_{s},N_{\tau})$. It turns out that these deviations occur because of the interplay between the tadpole and cactus diagrams in Fig.~\ref{self_energy_diag}.  
In particular, for fixed $g_{0}$ if $am_{0}$ is sufficiently small the cactus diagram contribution starts to compete with the tadpole diagram, despite being suppressed by a higher power in $g_{0}$. This behaviour stems from the fact that the free field propagator in Eq.~\eqref{free_prop} has a zero mode ($p=0$) contribution which increasingly diverges for smaller values of $am_{0}$. The deviations seen in the two-loop predictions are therefore ultimately a consequence of the vacuum-like pole structure of the free-field propagators used in the perturbative expansion\footnote{See Ref.~\cite{Lowdon:2024atn} for more details.}. This conclusion is consistent with the issues raised in Sec.~\ref{intro}, namely that finite-temperature perturbation theory breaks down if one chooses the basic field propagators to coincide with those of the vacuum theory.

\section{Spectral constraints}

\noindent 
In Refs.~\cite{Bros:1992ey,Buchholz:1993kp,Bros:1995he,Bros:1996mw} the authors set out to define a non-perturbative framework for describing scalar QFTs at finite temperature. A highly significant result of this framework is that the spectral function $\rho(\omega,\vec{p})$, defined as the Fourier transform of the thermal commutator $\langle [\phi(x),\phi(0)]\rangle_{\beta}$, must satisfy the following general representation:
\begin{align}
\rho(\omega,\vec{p}) = \int_{0}^{\infty} \! ds \int \! \frac{d^{3}\vec{u}}{(2\pi)^{2}} \ \epsilon(\omega) \, \delta\!\left(\omega^{2} - (\vec{p}-\vec{u})^{2} - s \right)\widetilde{D}_{\beta}(\vec{u},s).    
\label{BB_rep}
\end{align}
Equation~\eqref{BB_rep} is the finite-temperature generalisation of the well-known K\"{a}ll\'{e}n-Lehmann spectral representation that exists in vacuum QFTs~\cite{Kallen:1952zz,Lehmann:1954xi}. A major implication of Eq.~\eqref{BB_rep} is that the properties of $\rho(\omega,\vec{p})$ are entirely determined by the thermal spectral density $\widetilde{D}_{\beta}(\vec{u},s)$. Understanding the behaviour of $\widetilde{D}_{\beta}(\vec{u},s)$ is therefore crucial for establishing the type of scalar excitations that can exist at finite temperature. In Ref.~\cite{Bros:1992ey} it was further proposed that if a stable particle state of mass $m$ exists at zero temperature, then in position space the thermal spectral density $D_{\beta}(\vec{x},s)$ must contain a discrete component of the form: $D_{m,\beta}(\vec{x})\, \delta(s-m^{2})$. In the $T\rightarrow 0$ limit: $D_{m,\beta}(\vec{x}) \rightarrow 1$, and hence $\rho(\omega,\vec{p})$ reduces to the form for a vacuum particle state. But when $D_{m,\beta}(\vec{x})$ has a non-trivial structure this causes the zero-temperature peak in $\rho(\omega,\vec{p})$ to become broadened, which is a realisation of the dissipative effects that the vacuum particle state experiences as it moves through the medium. These discrete contributions therefore describe particle-like excitations of the thermal medium, and are referred to as \textit{thermoparticles}~\cite{Buchholz:1993kp}. \\

\begin{figure}[t!]
\centering
\includegraphics[width=0.49\textwidth]{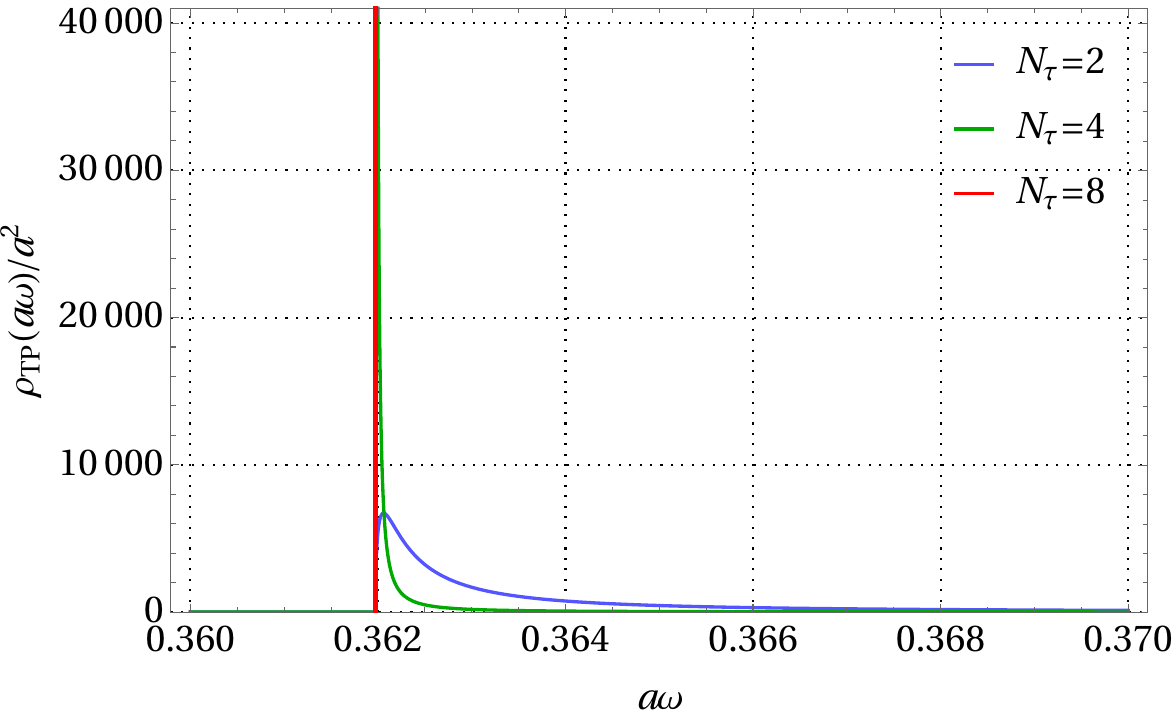} 
\includegraphics[width=0.46\textwidth]{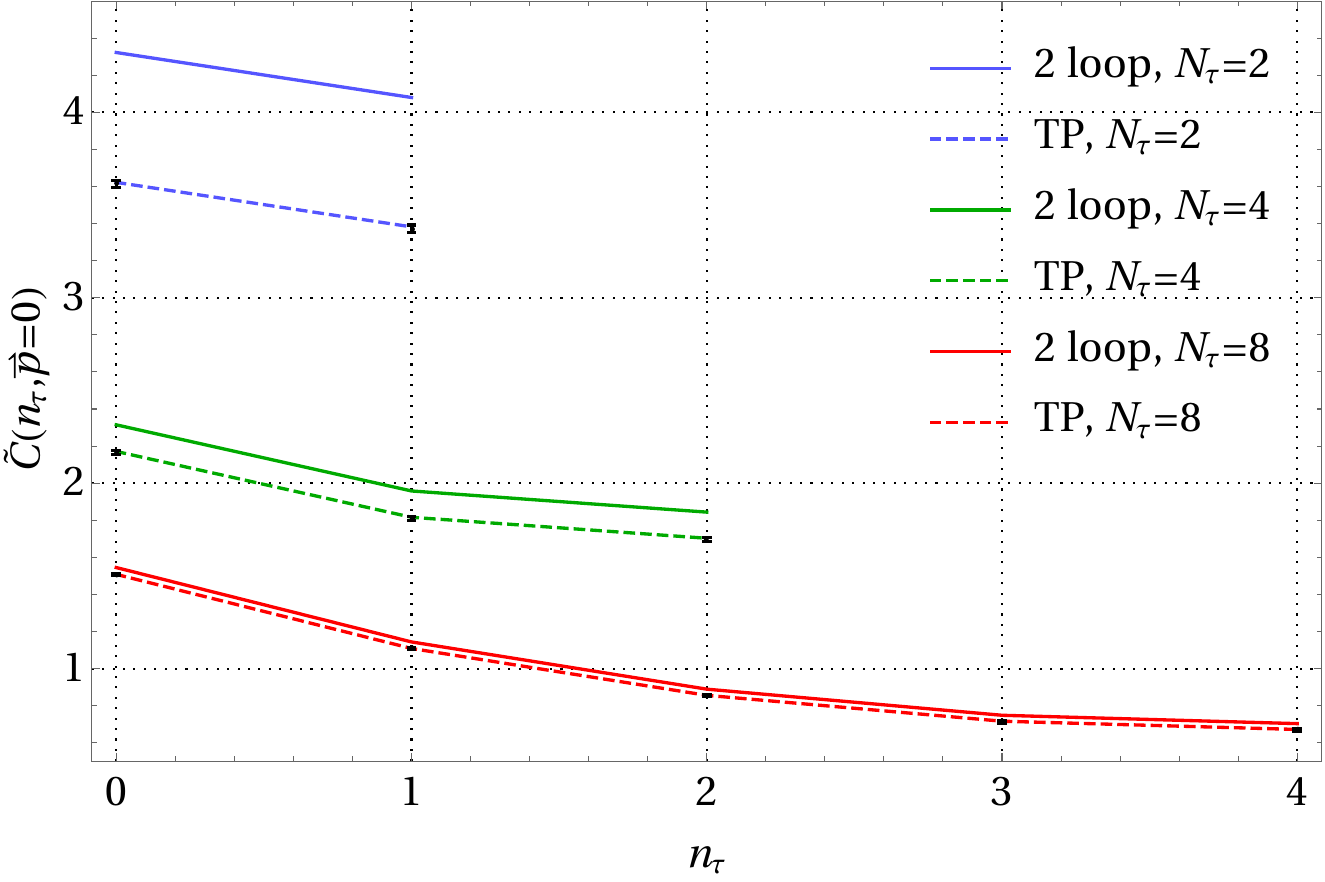}
\caption{Lattice temporal correlator data and predictions for $N_{s}=16$ and $(am_{0}=0.15,g_{0}=1.5)$ (right). The solid lines indicate the two-loop predictions from lattice perturbation theory, and the dashed lines are the predictions using a thermoparticle (TP) spectral function, the parameters of which are extracted from the spatial correlator data. The form of the corresponding TP spectral functions are displayed in the left plot for each temperature.}
\label{CtPred16}
\end{figure}

\noindent
In the next stage of the analysis in Ref.~\cite{Lowdon:2024atn} we investigated whether the lattice $\phi^{4}$ theory data is consistent with the presence of thermoparticle excitations. For this purpose we adopted the same approach as set out in Refs.~\cite{Lowdon:2022xcl,Bala:2023iqu}, where lattice QCD correlator data was used to establish that light pseudo-scalar mesons display thermoparticle-like behaviour around the chiral crossover region. This involved fitting the spatial correlator data to extract potential thermoparticle spectral components, and then using these components to predict the form of the temporal correlator, and comparing these predictions with the corresponding lattice data to check the consistency of the approach. More details of the specific procedure can be found in Ref.~\cite{Lowdon:2024atn}. The results from this analysis are displayed in Fig.~\ref{CtPred16}. The left plot of Fig.~\ref{CtPred16} shows the form of the zero-momentum thermoparticle spectral functions extracted from the spatial correlator data, and in the right plot we compare the temporal correlator predictions using these spectral functions (dashed lines) and the two-loop perturbative predictions (solid lines) with the corresponding lattice data. These findings show that unlike perturbation theory, the thermoparticle predictions are consistent with the lattice data at each temperature. This suggests that just like in QCD~\cite{Lowdon:2022xcl,Bala:2023iqu} the thermal medium of $\phi^{4}$ theory contains thermoparticle excitations which dominate at low temperatures. Although the thermoparticle predictions in Fig.~\ref{CtPred16} are purely non-perturbative, this indicates that these excitations could in fact provide the basis of a consistent finite-temperature perturbative expansion, as detailed in Ref.~\cite{Lowdon:2024atn}.  

\section{Conclusions and outlook}

\noindent
Finite-temperature QFTs are subject to non-perturbative constraints which have significant implications for the perturbative formulation of these theories. A particularly consequential constraint arises from the Narnhofer-Requardt-Thirring (NRT) theorem~\cite{Narnhofer:1983hp}, which implies that any perturbative expansion constructed using free field, or quasi-particle propagators with purely real poles, is inconsistent. In Ref.~\cite{Lowdon:2024atn} we set out to test this constraint by comparing the standard perturbative predictions of scalar correlation functions in $\phi^{4}$ theory with the results from lattice simulations. By computing spatial correlator predictions up to two-loop order in lattice perturbation theory, we found that these predictions deteriorated with increasing temperature, and that this was a direct result of using vacuum-like free-field propagators in the perturbative expansion. These results are consistent with the conclusion of Ref.~\cite{Weldon:2001vt} that finite-temperature perturbative theory explicitly breaks down in $\phi^{4}$ theory at two-loop order. In the remainder of Ref.~\cite{Lowdon:2024atn} we investigated whether these issues can be resolved using the non-perturbative framework put forward in Refs.~\cite{Bros:1992ey,Buchholz:1993kp,Bros:1995he,Bros:1996mw}. An important consequence of this framework is that the thermal medium potentially contains distinct particle-like excitations, so-called thermoparticles. By analysing the lattice $\phi^{4}$ data we find evidence for the existence of such excitations, which suggests that a consistent perturbative expansion should be parametrised in terms of these non-perturbative degrees of freedom.

\section*{Acknowledgements}

\noindent
The authors acknowledge support by the Deutsche Forschungsgemeinschaft (DFG, German Research Foundation) through the Collaborative Research Center CRC-TR 211 ``Strong-interaction matter under extreme conditions'' -- Project No. 315477589-TRR 211. O.~P.~also acknowledges support by the State of Hesse within the Research Cluster ELEMENTS (Project ID 500/10.006).

\bibliographystyle{JHEP}  
\bibliography{refs}

\end{document}